# Intermolecular dehalogenation reactions on passivated germanium(001)

*Nina C. Berner[1,2] \*, Yulia N. Sergeeva[3], Natalia N. Sergeeva[3,5], Mathias O. Senge[3], Attilio A. Cafolla[4] and Ignatius T. McGovern[1]*

[1] School of Physics, Trinity College Dublin, Dublin 2, Ireland
[2] CRANN, Trinity College Dublin, Dublin 2, Ireland
[3] School of Chemistry, SFI Tetrapyrrole Laboratory, Trinity Biomedical Sciences Institute, Trinity College Dublin, Dublin 2, Ireland
[4] School of Physical Sciences, Dublin City University, Glasnevin, Dublin 9, Ireland.
[5] School of Chemistry, University of Leeds, Leeds LS2 9JT, UK

ABSTRACT

We present results of experiments to reproduce the bottom-up formation of covalently bonded molecular nanostructures from single molecular building blocks, previously demonstrated on various coinage metal surfaces, on a technologically more relevant semiconductor surface: Ge(001). Chlorine was established as the most stable passivation agent for this surface, successfully enabling diffusion of the organic molecular building blocks. Subsequent thermal activation of the intermolecular dehalogenation reactions on this surface resulted in the desired covalently connected molecules, however showing poor network quality when compared to those formed on noble metal substrates.

INTRODUCTION

The miniaturization trend in the construction of electronic devices has recently sparked an interest in the bottom-up construction of two-dimensional nanostructures from single organic molecular building blocks on surfaces[1-3]. Covalently bonded networks are the most desirable due to their high thermal, mechanical and chemical stability[4-6]. The first covalently bonded nanoporous network was demonstrated by Grill and coworkers *via* a dehalogenation reaction of bromine-substituted tetraphenylporphyrin molecules on a Au(111) surface in ultra high vacuum (UHV)[7]. Dehalogenation reactions have subsequently been observed on other coinage metal substrates using a variety of different organic molecules with bromine or iodine end-groups[8-13]. Although the dehalogenation reaction is often referred to as an Ullmann-type reaction, which requires a catalyzing substrate[14], it has also been demonstrated on a bulk insulator surface[15] where a higher activation temperature is required.

This study explores the possibility of creating two-dimensional covalently bonded nanosized networks *via* intermolecular dehalogenation reactions on Ge, a technologically more relevant semiconductor surface. Silicon(001), the most common inorganic semiconductor surface used in devices today, may be the most obvious choice as a substrate. However, for initial investigations, germanium(001) has been chosen due to its higher tolerance against residual water contamination in UHV.

The molecular precursor used in this study is 5,10,15,20-tetrakis(4-bromophenyl)porphyrin (TBr$_4$PP, **1,** Figure 1), which was also used by Grill and coworkers in their pioneer work on on-surface intermolecular dehalogenation reactions, and 4',4''-diiodo-*p*-terphenyl (DITP, **2,** Figure 4). A schematic representation of the dehalogenation reaction between TBr$_4$PP molecules is shown in Figure 1.

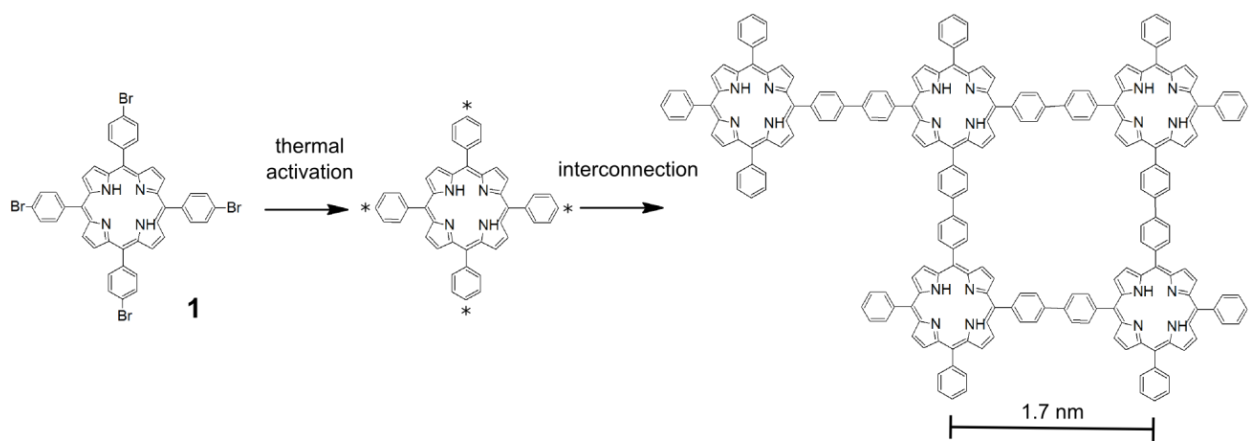

**Figure 1.** Schematic representation of the dehalogenation reaction between TBr$_4$PP molecules (**1**) into two-dimensional nanomesh networks.

A previous adsorption study of TBr$_4$PP on Ge(001) showed strong interactions between the clean substrate surface and both the molecular macrocycle and the bromine substituents[16]. Surface passivation is therefore required to inhibit these interactions and thus facilitate the molecular mobility required for intermolecular bond formation. However, the thermal activation (*i.e.* Br–C bond cleavage) of TBr$_4$PP on hydrogen-passivated Ge(001) has been observed to result in the premature desorption of the hydrogen from the surface, most likely due to HBr formation[16]. As part of this study, chlorine has been investigated as an alternative passivation agent for Ge(001) in UHV. It has a significantly stronger bond with germanium than hydrogen (349 kJ/mol *versus* 288 kJ/mol[17]) and has been established previously as a passivation agent for Ge(001) using wet-chemical methods[18].

EXPERIMENTAL METHODS

All experiments were performed under UHV conditions with a base pressure of less than $3\times10^{-10}$ mbar. Atomically clean surfaces of n-type Ge(001) were prepared by 20 to 40 minutes of Ar$^+$ sputtering with an ion energy of 500 eV and subsequent annealing at 850 K for 40 minutes. This cleaning cycle was repeated as necessary until an atomically clean Ge(001) surface was obtained, as indicated by low energy electron diffraction (LEED) and scanning tunneling microscopy (STM). After annealing the sample was cooled down slowly at a rate of 10 K/min until 600 K, after which it was cooled down more rapidly. This procedure resulted in a mixture of 2×1, c(4×2) and p(2×2) reconstructed domains on the Ge(001) surface.

Passivation of the Ge(001) surface with chlorine in UHV was achieved by deposition from an homebuilt electrochemical halogen cell, based on the design developed by Spencer and coworkers[19]. This method allows the controlled and minimized dosage of the comparatively reactive chlorine gas, avoiding corrosion damage to vacuum hardware. The chlorine exposure in units of Langmuir (L) was estimated using the partial chlorine pressure as determined by a Thermo Vacuum SMART IQ$^+$ mass spectrometer.

The TBr$_4$PP molecules were synthesized following a published procedure[7]. They were first purified by degassing at a temperature of 450 K for several hours and then deposited onto the Cl-passivated Ge(001) surfaces by sublimation from a Knudsen cell; at a cell temperature of 550 K, the deposition rate was 1 ML/h or less, depending on the sample orientation and chamber geometry.

The DITP molecules used in a comparison experiment were purchased from Tokyo Chemical Industry Europe and required a Knudsen cell temperature of 420 K for evaporation at a similar flux to that of the TBr$_4$PP.

STM images were obtained with an Omicron VT STM system in constant current mode and at room temperature, using electrochemically etched tungsten tips. X-ray photoelectron spectra (XPS) were recorded using an Omicron XPS system comprised of a DAR 400 dual anode X-ray source (Mg/Al) and a five channeltron EA125 hemispherical electron energy analyzer.

RESULTS AND DISCUSSION

The *in situ* passivation of Ge(001) with chlorine in UHV was investigated using STM and XPS. Figure 2a shows a filled state STM image of the chlorine-saturated Ge(001) surface, which was obtained after exposing the surface to 1 L of chlorine. This shows that the sticking coefficient of chlorine on Ge(001) is roughly unity, reflecting its comparatively high reactivity and strong bond to germanium. The STM image shows that, similar to the hydrogen-passivation, each germanium dangling bond is saturated with a single chlorine atom, resulting in symmetric dimers. According to DFT calculations of bromine and iodine on Ge(001)[20,21], the halogen atoms, unlike hydrogen atoms[22,23], appear brighter than the surrounding germanium atoms in filled state STM images, leading to the interpretation of the observed dark features as defects in the passivation, *i.e.* missing chlorine atoms (marked "MC" in Figure 2a). Bright features in the STM images are interpreted as a dichloride species (marked "DC" in Figure 2a). The average defect density of the chlorine-passivation has been determined to be 5 ± 2 %, which is only slightly higher than the lowest defect density achieved for the hydrogen-passivated Ge(001) surface of less than 1 % [23].

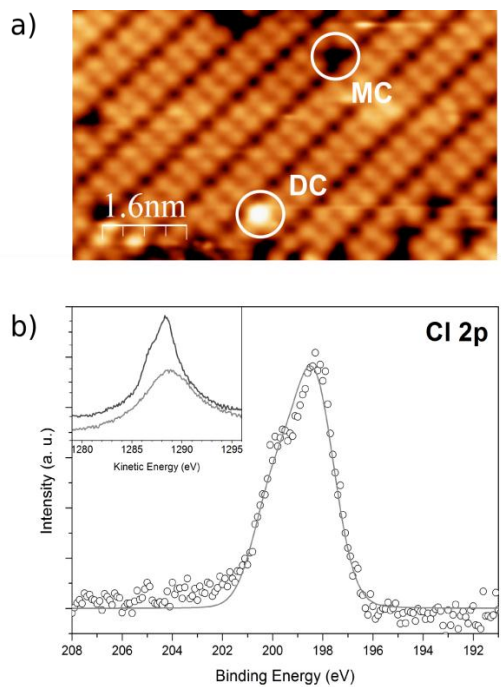

**Figure 2.** Chlorine-passivated Ge(001). a) STM image taken at -1.5 V and 0.23 nA, circles indicate a missing chlorine atom ("MC") and a possible dichloride ("DC"). b) XPS spectrum of the Cl 2p core-level (fitting parameters: Binding energy 198.4 eV, spin-orbit splitting 1.6 eV, branching ratio 0.5, Gaussian width 1.85 eV, Lorentzian width 0.15 eV), inset: Raw data before (light grey) and after (dark grey) chlorine-passivation.

XPS spectra were recorded to confirm chlorine-saturation of the surface. The Cl 2s and Cl 2p core-levels are difficult to measure quantitatively with both the Mg and the Al characteristic $K_\alpha$ x-rays since they have an energy overlap with either germanium Auger lines or plasmon lines, a problem which was previously identified by Ardalan and coworkers[24]. Figure 2b shows a spectrum of the Cl 2p core-level obtained by subtraction of the Ge 3s plasmon loss peak recorded on the clean surface from that of the Cl 2p doublet recorded from the chlorine-passivated surface. The inset shows the raw spectra of the clean surface (light grey) and after chlorine-passivation (dark grey). After removal of the increased scattering background on the high binding energy shoulder of the difference spectrum, the doublet peak could be fitted reasonably well using a spin-orbit split doublet with a Voigt line-shape. The binding energy of 198.4 eV is in good agreement with the 198.7 eV reported by Ardalan and coworkers for the Ge(001) surface which was chlorine passivated using a wet-chemically technique[24]. A systematic desorption study of chlorine on germanium has not been performed; however, STM images taken after several annealing steps indicate that Cl desorption is initiated between 520 and 620 K. Figure 3a shows a filled state STM image recorded at room temperature after the deposition of a sub-monolayer coverage of TBr$_4$PP on the Ge(001):Cl interface. The streaky nature of the image indicates molecular mobility on the surface and thereby confirms the successful suppression of the strong molecule-substrate interactions as observed for adsorption of TBr$_4$PP on the unpassivated

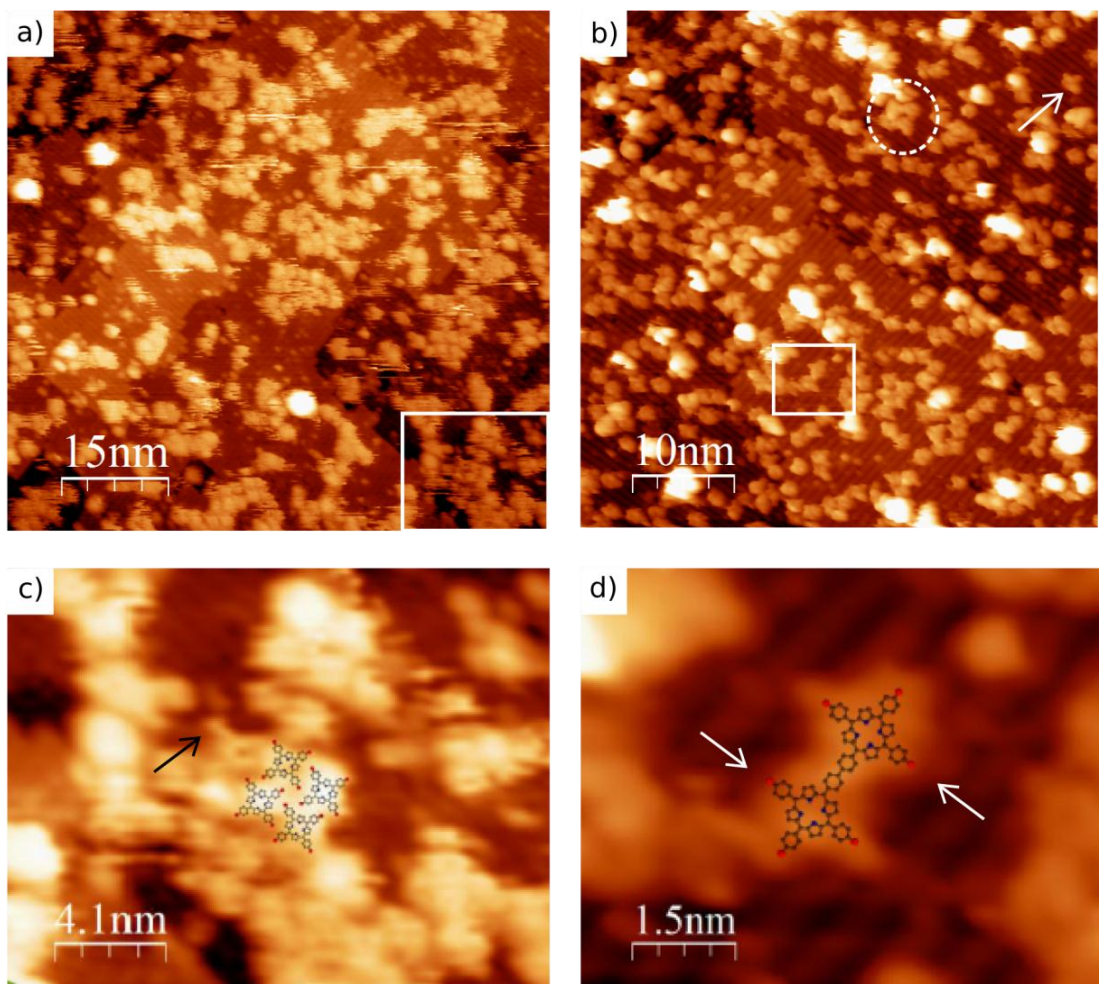

**Figure 3.** STM images of TBr$_4$PP on Ge(001):Cl before (right column) and after (left column) annealing at 520 K. a) was taken at -1.7 V and 0.09 nA, b) was taken at -1.9 V and 0.11 nA, and c) and d) show close-ups of the white rectangles marked in a) and b), respectively; with molecular structure overlays. The circle in b) highlights a pore structure, the arrow an isolated but resolved molecule. The arrow in c) points at a resolved phenyl leg, and the arrows in d) highlight phenyl legs with missing bromine substituents.

Ge(001) surface. Despite the increased surface diffusion, it is possible to resolve individual TBr$_4$PP molecules. The self-assembly of the molecules into islands through non-covalent interactions is also observed. Figure 3c shows a close-up of one of the islands, revealing molecular resolution down to the phenyl substituents and, in some cases, the free base core of the macrocycle. The exact configuration of the self-assembled molecules is difficult to extract from the images, but is assumed to be similar to the T-configuration observed in the self-assembly of TPP molecules on other substrates[7,25], as indicated by the overlaid molecular structures.

Annealing at 520 K for 5 minutes changes the behavior of the molecules on the surface significantly, as can be seen in the filled state STM image in Figure 3b. None of the streaky features indicating mobile molecules can be observed and the self-assembled islands of molecules have mostly transformed into chain-like structures. Images recorded with molecular resolution show strong evidence for leg-to-leg orientation of adjacent molecules without dark contrast in between, which, in conjunction with the significantly reduced surface mobility, indicates the successful debromination of the molecules and subsequent formation of covalent intermolecular bonds. Additionally, the average distance between molecular cores in the dimers, chains and pores observed throughout the STM images has been determined as $d = 1.65 \pm 0.12$ nm, which is in agreement with the $d = 1.71$ nm measured and calculated by Grill and coworkers for the distance between two covalently bonded TPP molecules. The close-up of a dimer shown in Figure 3d is sufficiently resolved to distinguish between activated and intact phenyl legs (with and without bromine substituent, respectively). Attempts to image this surface at higher biases to further confirm the covalent nature of the intermolecular bonds as previously demonstrated by Grill and coworkers resulted in cleavage of Ge–Cl bonds on the passivated surface and a strong disturbance of the STM images.

The size and quality of the observed covalently bonded network fragments is poor when compared to the results achieved on coinage metal surfaces[8-13]. The presence of single, isolated molecules which can be resolved with the STM, as indicated by the arrow in Figure 3c, indicates that there are anchor points on the substrate, holding the molecules in place. These are most likely the unavoidable defects, shown in Figure 2, *i.e.* remaining dangling bonds, in the surface passivation layer, which limit the mobility of the precursor molecules and therefore the quality of the resulting networks.

XPS data is required to determine whether the activation, *i.e.* dehalogenation of the TBr$_4$PP molecules on the Ge(001):Cl substrate, leads to the premature desorption of the passivation layer and subsequent formation of Br–Ge bonds as previously observed on Ge(001):H. However, the Br 3d core level has a very low photoionization cross-section for the photon energies provided by the Al and Mg anodes[26], and therefore does not yield sufficient intensity to analyze the spectra for sub-monolayer coverages of TBr$_4$PP. In contrast, the cross-section of the I 3d core-level is an order of magnitude higher than that of the Br 3d core level, which motivated a comparison experiment using the iodine-substituted 4',4"-diiodo-*p*-terphenyl (DITP) molecule as a precursor for an analogous dehalogenation reaction on Ge(001):Cl.

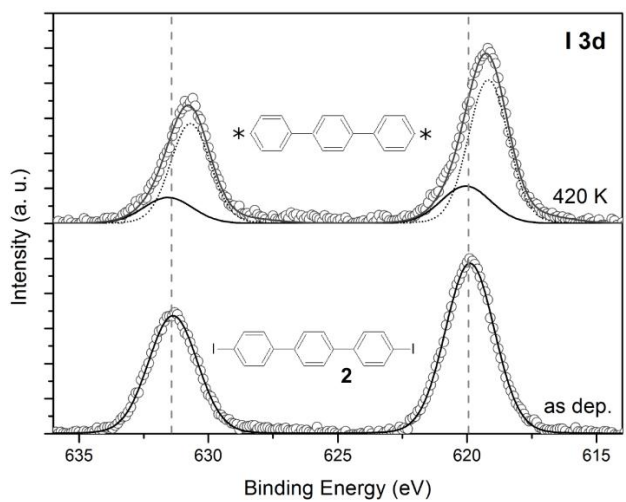

**Figure 4.** XPS spectra of the I 3d core-level of DITP (**2**) on Ge(001):Cl, before and after annealing at 420 K. The lower spectrum shows only one component at a binding energy of 620 eV, whereas this component is reduced to ~20 % of its original intensity after annealing (as shown in the upper spectrum), and a second component can be observed at 619.1 eV. For both fitting procedures, a spin-orbit splitting of 11.5 eV, a branching ratio of 0.66, a Gaussian width of 2 ( ± 0.2 ) eV and a Lorentzian width of 0.15 eV have been used.

The lower spectrum in Figure 4 shows the spin-orbit split I 3d core-level recorded after the deposition of a sub-monolayer coverage of DITP. The the 3d$_{3/2}$ and 3d$_{5/2}$ core levels at binding energies of 631.5 and 620.0 eV respectively are each fitted with a single Voigt component and result from iodine attached to the molecules[13]. After annealing the system at 420 K, the I 3d$_{3/2}$ and 3d$_{5/2}$ core-levels show two components shifted to lower binding energy by 0.9 eV as shown in the upper plot in Figure 4. The original 3d$_{3/2}$ component at 620 eV is reduced to ~20 % of its original value, indicating the successful dehalogenation of a substantial fraction of the adsorbed molecules. The second component at 619.1 eV is associated with iodine bonded to germanium, which indicates that the chlorine-passivation is partially destroyed in the annealing process. At a temperature as low as 420 K, this can only be explained by Cl–I formation, analogous to that reported for intermolecular dehalogenation reactions on the hydrogen-passivated substrate. Considering the low activation temperatures for dehalogenation of the molecules on Ge(001):Cl when compared to the insulator substrate (iodine activation at 530 K [15]) and inside the Knudsen cell (bromine activation at 590 K [7]), it is possible that the Cl–X and Ge–X formation serve as catalyzing side-reactions (where "X" represents the halogen substituent cleaved from the molecule).

STM of the DITP molecules on Ge(001):Cl before and after annealing repeatedly did not resolve any individual molecules or self-assembled structures; most likely due to interactions between the molecules and the STM tip that are stronger than the molecule-substrate interactions. We also observed this behavior in STM experiments with other organic molecules which are smaller than porphyrins and lack a planar macrocycle. It appears that the non-covalent molecule-substrate interactions are significantly enhanced by the presence of an extended π-orbital system of a macrocycle as in TBr$_4$PP, which makes it possible for macrocyclic molecules on Ge(001):Cl to be resolved in STM images.

CONCLUSIONS

After establishing a methodology for the Cl-passivation of the Ge(001) surface in UHV, STM images of TBr$_4$PP molecules after a thermally induced dehalogenation reaction show evidence for the formation of covalent bonds between molecules. However, the size and quality of the networks are very poor when compared to the results achieved with the same molecular precursors on noble metal surfaces. This is most likely caused by the intrinsic and unavoidable defects in the passivation layer, which limit the mobility of the molecules and therefore efficient network formation. Furthermore, it is suggested that the dehalogenation of the molecules induces the premature partial desorption of the underlying passivation layer, leading to more defects and making the dehalogenation reactions on passivated Ge(001) a self-limiting process.

AUTHOR INFORMATION

**Corresponding Author**

* Email: nberner@tcd.ie, phone: +35318964628

ACKNOWLEDGMENT

This work was supported by Science Foundation Ireland, through Principal Investigator Grants (grant numbers 09/IN.1/I2635 and 09/IN.1/B2650). NCB thanks Trinity College Dublin for a Postgraduate Award. STM topographic images were processed using WSxM software[27].

TABLE OF CONTENTS GRAPHIC

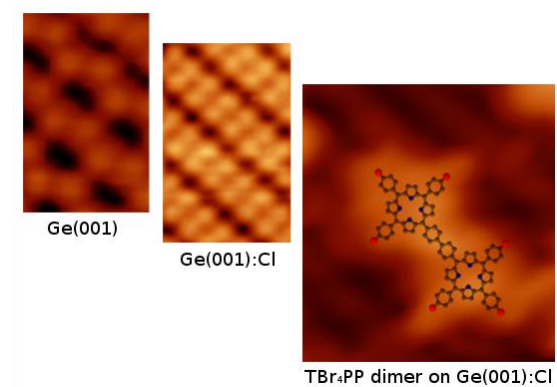